\documentclass[aps]{revtex4}
\usepackage{graphicx,amssymb}

\begin{document}
\title{Charge Fluctuations at Disoriented Chiral Transition}
\author{Rudolph C. Hwa$^1$ and C.\,B.\,Yang$^{1,2}$}
\affiliation{$^1$Institute of Theoretical Science and Department of
Physics, University of Oregon, Eugene, OR 97403-5203, USA\\
$^2$Institute of Particle Physics, Hua-Zhong Normal
University, Wuhan 430079,\\ P.\ R.\ China}
\date{October 2001}

\begin{abstract}
The signatures of the creation of a disoriented chiral condensate
 in heavy-ion collisions are studied in the Ginzburg-Landau
description of  chiral phase transition. Scaling properties are
found that characterize the fluctuations  of charged particles.
There exists also an explicitly calculable measure that can
directly be measured. Experimental procedures needed to find the
signatures are discussed.

\end{abstract}


The subject of event-by-event fluctuations of
charged particles produced in high-energy heavy-ion
collisions has recently gained considerable attention
\cite{jk,ahm,hj,gpz,rev}.  The focus has mainly been in the
identification of a measure that can differentiate a
quark-gluon plasma (QGP) from a hadron gas (HG).  No
consideration has been given to the fluctuations introduced by
the phase transition (PT) itself.  If the PT corresponds to the
second-order chiral transition that has the usual $O(4)$
symmetry of two massless quarks, it has been argued that
long-range correlation typical of critical phenomena is
unlikely to occur because the only scale in the problem is
characterized by the realistic pion mass that is roughly the
same as the critical temperature \cite{raj}. As a result one does
not expect large fluctuations, especially in the charge sector. On
the other hand, if the chiral transition takes place rapidly far
from thermal equilibrium, then the disorientation of the
isovector order parameter can lead to long-wavelength modes
and large fluctuations \cite{rw,bk,hw}.  The usual signature for
such disoriented chiral condensates (DCC) is the large
fluctuation in the neutral-to-charge ratio of the produced
particles
\cite{and,ans,kt,bj}. In this paper we study the nature of the
fluctuations within the charge sector and identify measures
that can be computed explicitly as well as being amenable to
direct experimental verification. They do not depend on the existence of
long-range correlations that may be suppressed by the rapid expansion in
a heavy-ion collision.

The usual starting point of a consideration of DCC is the
linear $\sigma$ model, for which the potential is
\begin{eqnarray}
V={\lambda\over 4}(\Phi^2-v^2)^2 - H\sigma,
\label{1}
\end{eqnarray}
where the chiral fields are represented by a vector
$\Phi=(\sigma, {\vec \pi})$ in $O(4)$ space. The parameters
$\lambda, v,$ and $H$ are determined by the masses $m_{\pi},
m_{\sigma}$ and the pion decay constant $f_{\pi}$. At
temperature $T$ far below the critical $T_c$, the normal
vacuum is characterized by $\left<\sigma\right>\neq 0,$ and
$\left<\vec\pi\right>=0$. At $T$ well above $T_c$ there is
approximate $O(4)$ symmetry for which $\left<\Phi\right>=0$.
If the QCD plasma is close to thermal equilibrium as it cools
from above $T_c$ to below, the PT results in
$\left<\Phi\right>$ becoming nonzero in the $\sigma$ direction
with no large fluctuations expected in the angular deviation
from $\sigma$. However, in the quench scenario far from
thermal equilibrium \cite{rw}, the plasma loses touch with the
vacuum orientation and $\left<\vec \pi\right>$ becomes
nonzero along arbitrary directions in isospin space in different
spatial regions, thereby generating large fluctuations in the
charges of the pions produced. What we seek are the
signatures of those fluctuations that are independent of the
details of the theory, specifically, the parameters governing
the chiral transition.
We show the existence of a numerical
index $\nu$ that can serve as a signature of DCC. As an
alternative to the $D$ measure that has been suggested in
\cite{jk,gpz}, we consider another measure $B$ that can clearly
distinguish the different types of charge fluctuations.

It is convenient to start with the coherent-state
representation for the statistical fluctuations since the
multiplicity distribution of a pure coherent state
$|\alpha\rangle$ is Poissonian, i.e., $\left|\left<n|\alpha
\right>\right|^2 = P^0_n$, with average multiplicity
\begin{eqnarray}
\left<n \right> = \left<\alpha
\left|\int dz a^{\dagger}(z) a(z)\right|\alpha \right> = \int
dz \left| \alpha(z)\right|^2  ,
\label{2}
\end{eqnarray}
where the property $\left. a(z)|\alpha \right> = \left.
\alpha(z)|\alpha \right>$ has been used.  We generalize this
formalism by incorporating isospin and treat ${\vec\phi}$
as  the eigenvalues of the isovector annihilation operators
${\vec a}(z)$  \cite{ss}.  The total average density
of hadrons (assumed to be pions only) is then
\begin{eqnarray}
 \left<{\vec \phi} \left| {\vec a}^{\dagger}(z) {\vec
a}(z)\right|{\vec\phi}\right> = \left| {\vec\phi}(z)\right|^2 .
\label{3}
\end{eqnarray}
The spatial coordinate $z$ can be regarded as (pseudo) rapidity
in heavy-ion collisions, as in \cite{ss}, but at this point that
identification is unnecessary. Our theoretical results, expressed below by Eqs. 
(16), (17) and (21), are independent of what $z$ is exactly. For experimental 
analysis of the data, extensive discussion will be given below.

The application of coherent states and the Ginzburg-Landau
formalism \cite{gl} to multiparticle production was considered
many years ago \cite{ss}. We use the same approach here, as in
\cite{hn,hp}, to study the thermal fluctuations in the chiral
transition to DCC. We take the Ginzburg-Landau free energy to
be
\begin{eqnarray}
F[{\vec \phi}] = \int_{\delta} dz \left[ a\left| {\vec \phi}(z)
\right|^2 + b\left| {\vec\phi}(z)
\right|^4 \right] ,
\label{4}
\end{eqnarray}
where only the isosymmetric part of the potential  in
Eq.\,(\ref{1}) is adapted here. The derivative term in (4) is neglected here, 
since our earlier studies of isoscalar $\phi(z)$ indicate that the inclusion of
$\partial\phi (z) /\partial z$ term in the free energy leads to negligible effect
on the scaling result \cite{hp,yc}. When $T$ is lowered below $T_c$,
$a$ becomes negative, while $b$  remains positive,
and the system makes a transition to the hadron phase
whose density fluctuates around $\left|{\vec
\phi}\right|^2 = -a/2b$.  The hadronic multiplicity
distribution is then given by
\begin{eqnarray}
\label{5}
P(n_+, n_-, n_0) = Z^{-1} \int \mathcal{D} {\vec \phi}\ P^0(n_+, n_-,
n_0, \left| {\vec \phi}\right|^2) e^{-F[{\vec\phi}]}
\end{eqnarray}
where
\begin{eqnarray}
\label{6}
Z = \int \mathcal{D} {\vec \phi}\ e^{-F[{\vec\phi}]}   ,
\end{eqnarray}
\begin{eqnarray}
\label{7}
P^0(n_+, n_-, n_0, \left| {\vec\phi}\right|^2) = \prod_{i =
+,-,0} P^0(n_i, \left| \phi_i \right|^2)  ,
\end{eqnarray}
\begin{eqnarray}
\label{8}
P^0(n_i, \left| \phi_i \right|^2) = { 1 \over  n_i!} \left(\int_\delta
dz  \left| \phi_i \right|^2\right)^{n_i} e^{-\int_\delta
dz  \left| \phi_i \right|^2}  .
\end{eqnarray}
Since these quantities are meaningful only in the hadron
phase, we shall in the following consider only the  situation
where $a<0$.

The central aim of our proposed analysis is to find a
measure of the charge fluctuations due to a chiral transition to
arbitrary $\left<\vec\pi\right>$ directions with the property
that the measure is independent of the details, more
specifically, the parameters
$a$, $b$, and $\delta$ in Eq.\,(\ref{4}).   We shall assume that
$\delta$ can be arranged to be small so that $\vec\phi$ can be
regarded as constant inside $\delta$. We then have
\begin{eqnarray}
F[\vec\phi] = \delta \left[a\left| {\vec\phi}\right|^2 + b\left|
{\vec\phi}\right|^4\right]  ,
\label{9}
\end{eqnarray}
where the  value of each of $\left| {\it \phi}_i\right|$
 is allowed to vary throughout the complex plane in
the functional integrals in Eqs.\,(\ref{5}) and (\ref{6}).

In our search for a quantity that is independent of
$\delta$, $a$, and $b$, we first consider the bivariate
factorial moments
\begin{eqnarray}
f_{q_1, q_2} = \sum^{\infty}_{n_+=q_1}\sum_{n_-=q_2}^\infty
\sum_{n_0=0}^{\infty}{n_+!  \over  \left(n_+ - q_1\right)!}{n_-!\over
\left(n_- - q_2\right)!} P (n_+, n_-, n_0)  ,
\label{10}
\end{eqnarray}

\noindent for integer values of $q_1$ and $q_2$.
If $P (n_+, n_-, n_0)$ were the statistical distribution given
in Eqs.\,(\ref{7}) and (\ref{8}), then (\ref{10}) would yield
$f_{q_1, q_2} =f_{0,1}^{q_1}f_{1,0}^{q_2}$
for any $q_1$ and $q_2$.  Deviation
from this trivial result on account of Eqs.\,(\ref{5}) and
(\ref{8}) is then a measure of the effect of the DCC
 on the charge fluctuations plus other effects to be discussed below.
Making the appropriate substitutions we obtain
\begin{eqnarray}
f_{q_1, q_2} &=&Z^{-1} \delta^{q_1+ q_2} \,\pi^3
\,\int^{\infty}_{0}d\left|\phi_+\right|^2
 \, \int^{\infty}_{0}d \left| \phi_-\right|^2 \,
\int^{\infty}_{0}d \left| \phi_0\right|^2 \nonumber\\
&&
\left|\phi_+\right|^{2q_1}\, \left|\phi_-\right|^{2q_2}
e ^{- \delta \left(a\left| {\vec\phi}\right|^2 + b\left|
{\vec\phi}\right|^4 \right)}  .
\label{11}
\end{eqnarray}
Changing the integration variables to the set $s=|\phi_+|^2,
t=|\phi_+|^2+|\phi_-|^2$, and $u=|\vec\phi|^2$, so that only
$u$ is integrated from 0 to $\infty$, we obtain a significantly
simplified, closed form
\begin{eqnarray}
f_{q_1, q_2} = \left({\delta  \over  b}\right)^{\left(q_1+
q_2 \right)/2} \, {2B \left(q_1 + 1, q_2  +1\right)  \over
q_1 + q_2 +2} \, {J_{q_1 + q_2 +2}(x)  \over  J_2 (x)} ,
\label{12}
\end{eqnarray}
where $B(m,n)$ is the Euler-beta function, and
\begin{eqnarray}
x = \left| a \right| \sqrt{\delta/b} .
\label{13}
\end{eqnarray}
The function $J_p(x)$, defined by
\begin{eqnarray}
J_p(x) = \int^{\infty}_{0}d u \, u^p\, e^{xu-u^2}  ,
\label{14}
\end{eqnarray}
can be related to the parabolic cylinder function, but is
straightforwardly computable.

Although $f_{q_1, q_2}$ has complicated dependence on
$\delta$, $a$ and $b$, different $(q_1, q_2)$ moments
have the same dependence if $q_1+ q_2 \equiv q$ is the
same.  Furthermore, the factor $(\delta/b)^{(q_1+
q_2)/2}$  is cancelled for the normalized factorial
moments, which we define as
\begin{eqnarray}
&& F_{q_1, q_2}\equiv {f_{q_1, q_2}\over f_{1,0}^{q_1}
f_{0,1}^{q_2}}\nonumber\\
&=& {2B \left(q_1 + 1, q_2  +1\right)  \over
q_1 + q_2 +2} \, \left[ {3J_2(x)  \over  J_3
(x)} \right]^{q_1 + q_2 }\, {J_{q_1 + q_2 +2}(x)  \over  J_2
(x)}  .
\label{15}
\end{eqnarray}
Evidently, $F_{q_1, q_2}(x)$ is a function of $x$ only.  Its
dependence on $x$ is shown in Fig.\,1 in a log-log plot.  No
scaling behavior can be seen.  However, if we plot $\ln F_{q_1, q_2}$ vs
$\ln F_{2, 2}$ as in Fig.\,2, we find a
substantial region in which the relationship is linear.  In
that linear region we can write
\begin{eqnarray}
F_{q_1, q_2} \propto F_{2, 2}^{\beta_{q_1, q_2}}  ,
\label{16}
\end{eqnarray}
which is a behavior that is independent of $x$.  More
explicitly, we can determine $\beta_{q_1, q_2}$ by
straightline fits of the curves in Fig.\,2 in the region $0.2
< \ln F_{2, 2} < 0.9$.  The result yields $\beta_{q_1,
q_2}$ as a function of only the sum, $q_1 + q_2 = q$, as
can be seen directly from Eq.\ (\ref{15}).  That dependence
of $\beta_q$ on $q$ is shown in Fig.\,3.  Apart from the point
for $q = 2$, it is a linear dependence of $\ln \beta_q$
 on $\ln (q - 1)$.  Thus we can write
\begin{eqnarray}
\beta_q \propto (q - 1)^{\nu}  , \hspace{1cm} \nu = 1.29 .
\label{17}
\end{eqnarray}
This index $\nu$ is independent of $x$, and therefore of
$\delta$, $a$ and $b$ (so long as $a < 0$).
The scaling behaviors (\ref{16}) and (\ref{17}) summarize
the properties of charge fluctuations in a chiral transition to
DCC and culminate in a numerical index $\nu$ that
characterizes the   phenomenon.

To verify the above behavior experimentally, it is necessary to
vary $x$, which is the implicit variable in Eq.\ (\ref{16}). Since
$a$ and $b$ in Eq.\ (\ref{13}) are not subject to experimental
control, only $\delta$ can be varied. A
more extensive discussion of the experimental cuts that are
optimal for detecting the signal is postponed until another
prediction independent of $x$ is presented.

A few remarks should first be made regarding the sensitivity of the
above theoretical result to statistical fluctuations in the experimental
background. It was shown by Bia{\l}as and Peschanski \cite{bp} that
the factorial moments filter out the statistical fluctuations represented
by the Poissonian distribution in Eq.(\ref{8}). However, if the DCC is
produced in a background of conventionally produced hadrons, we
have to consider an additional contribution to the mean multiplicity.
Thus we replace $|\phi_i|^2$ on the RHS of (\ref{8}) by $|\phi_i|^2 + S$,
where $S$ denotes the mean density of the statistical background. If there are
statistical fluctuations in the orientations of the fields after the chiral transition,
we shall let $S$ represent them also. We
ask whether our result is sensitive to a small perturbation by $S$.
Clearly, if $S$ were large, it would be hard  to find the signature of
DCC in the presence of a large non-DCC hadronization process. If $S$ is
small, we can carry the above theoretical calculation to first order in
$S$, and find that a term must be added to $F_{q_1,q_2}$ in (\ref{15}),
which is proportional to $S\sqrt {b\delta}$. At small $\delta$ the effect
of $S$ is expected to be negligible.

There is, however, a limitation on how small $\delta$ should be allowed to
be. That can be seen from Eq.(\ref{11}) where the exponential damping term
becomes ineffective at small $\delta$. Without letting $|\vec\phi|^2$
become large enough to require higher-order terms in the Ginzburg-Landau
free energy, a rough lower bound on $\delta$ is derived in Ref. \cite{hn}
and is equally applicable here; it is $\delta>x_0^2b/a^2$, where
$x_0=\sqrt{4\ln 2}=1.67.$ That bound translates to
$S\sqrt{b\delta}>(Sb/|a|)x_0.$ Thus the sensitivity to the statistical
background depends on its strength relative to that of the dynamical
strength $|a|/b$, as is reasonable. Our conclusion is therefore that
unless the mean multiplicity of the statistical background is small
compared to that of the DCC production, our approach of searching for
scaling behavior would not be successful.

\indent
To put the curves in Fig.2 to experimental test, one inevitably has to
deal with the inaccuracies in the determination of $F_{q_1,q_2}$, and
in particular with $F_{2,2}$. If the error bars are large, the
effectiveness of this approach is clearly limited. Even if they are not
large, the data points may admit a larger scaling region than what we
have consider in Fig.2 in the derivation of $\nu=1.29$. In fact, our lower
bound on $\delta$ mentioned above implies that the curves in Fig.1 are
reliable only for $x<x_0$, or $-\ln x < -0.51$. It translates to a
restriction on the range of $F_{2,2}$ to the region $\ln F_{2,2} < 0.65$
in Fig. 2. Thus fitting the curves in that region by straight lines with
allowance for errors we find that it is necessary to modify the value of
$\nu$ to
\begin{equation}
\nu = 1.42\pm 0.13.     \label{17a}
\end{equation}
It is clear that we can no longer claim strict x-independence in our
result, when the various complications discussed above are taken into
account. However, despite the limitations arising from theoretical
considerations, we feel that the proposal for the experimental measurement
of $F_{q_1,q_2}$ may nevertheless yield interesting insight into the formation of DCC.

To reduce the sensitivity to the details in the scaling analysis, we now
consider a global measure that makes contact with the  observables
proposed in \cite{jk,ahm,hj,gpz,rev}. Let us first list the simple identities:
\begin{eqnarray}
f_{1,0} = \left<n_+ \right>,\quad f_{0,1} = \left<n_- \right>,\quad
f_{1,1} = \left<n_+ n_-\right>,\nonumber\\
 f_{2,0}= \left<n_+(n_+-1)\right>, \quad f_{0,2}
= \left<n_-(n_--1)\right>.
\label{18}
\end{eqnarray}
In terms of the usual definitions, $N_{\rm ch} = n_+ + n_-$,
$Q = n_+ - n_-$, and $\left<\delta X^2 \right> = \left<
X^2\right> - \left< X\right>^2$, we then have
\begin{eqnarray}
\left< \delta Q^2\right> = 2 \left(f_{2,0} + f_{1,0} - f_{1,1}
\right)  ,
\label{19}
\end{eqnarray}
where the symmetry $f_{i,j} = f_{j,i}$ has been used.  Let
us now consider the quantity
\begin{eqnarray}
B = {\left< \delta Q^2\right> - \left< N_{\rm ch}\right>
\over \left< n_+ n_-\right>}   .
\label{20}
\end{eqnarray}
Using Eqs.\ (\ref{18}) and (\ref{19}), together with
(\ref{12}) we obtain
\begin{eqnarray}
B = 2 \left({ f_{2,0} \over f_{1,1} }  -1\right) = 2 \left[{
B(3,1) \over B(2,2) }  -1\right]= 2  .
\label{21}
\end{eqnarray}
On the other hand, in terms of $D$ \cite{jk,gpz}, where
\begin{eqnarray}
D = 4 \left< \delta Q^2\right>/\left< N_{\rm ch}\right>  ,
\label{22}
\end{eqnarray}
Eq.\,(\ref{20}) implies
\begin{eqnarray}
B = {D/4 - 1  \over \left< n_+n_-\right>/\left<
N_{\rm ch}\right>}  .
\label{23}
\end{eqnarray}
It has been argued that $D \leq 4$ whether the thermal
system is a QGP or a HG \cite{jk}, so in that scenario we
have $B \leq 0$ whatever $\left< n_+n_-\right> / \left<
N_{\rm ch}\right>$ may be.  That prediction is distinctively
different from 2 in our scenario.  Qualitatively speaking,
the charge fluctuations in  chiral transition to DCC are
much greater than what can be expected from the thermal
fluctuations in either the QGP or the HG phases. Thus
the measurement of $B$ can decisively determine the nature of
the PT that the QGP system undergoes. The presence of statistical
fluctuations in the background will undoubtedly weaken this claim to an extent
that is proportional to the strength of such fluctuations, as we
have found in the study of the scaling behavior.

In order that our theoretical prediction can be applied to
the data, it is important that the data are analyzed in the
proper way, which is the subject we now address.  The
factorial moments can reveal scaling behavior if they can capture the rare
events with large spikes without being overwhelmed by the
contributions from average multiplicities \cite{bp}. In other words, if
we, for simplicity, consider a single factorial moment
\begin{eqnarray}
f_q = \sum^{\infty}_{n = q} n(n - 1) \cdots (n - q + 1) P_n ,
\label{24}
\end{eqnarray}
we see that $f_q$  probes the high end of the distribution
$P_n$ with $n\ge q$; thus if $q \gg\left<n\right>$, only  the
less frequent events with $n \gg \left< n\right>$ can
contribute to Eq.\,(\ref{24}).  That can be achieved for
heavy-ion collisions either by using extremely high $q$, or
by making severe cuts in data selection to reduce
$\left<n\right>$.  Neither was done in previous analyses of the
nuclear data and nothing of interest was found in $F_2$ or
$F_3$ in the bulk data.

To test the power-law behaviors (\ref{16})--(\ref{18})
and the prediction $B = 2$, it is necessary to make cuts in the
data to best reveal the features of this analysis. To see what
cuts to make, we first discuss multiparticle production in
heavy-ion collisions. If the quark-hadron PT is first-order,
then what we have considered here is irrelevant. For a
second-order chiral transition it is likely that, if the DCCs are to
be created at all, they would appear as clusters of hadrons in
different regions in space and time amidst other patches of
hadrons produced as a result of the gradual cooling of the QGP
that undergoes the normal critical transition. If it is possible to
restrict the observation to a very short duration in the
emission time, then we should see many regions of voids
where no particles are produced, separating the regions of
particles produced either as oriented or disoriented
condensates. Unfortunately, a selection in emission time is not
experimentally feasible. The event multiplicity integrates over
all times and smooths out the fluctuations in clusters and voids.
Thus to detect the DCC by our fluctuation analysis it is
necessary to make a severe cut in $p_T$ and select only the
particles emitted into a narrow $p_T$ window. Such a $\Delta
p_T$ cut achieves two goals:  one, to reduce the
multiplicities to facilitate the factorial moment analysis
discussed above; two, to minimize the overlap of hadron
emissions at different times.  The correlation between $p_T$ and
the emission time is discussed in Ref.\cite{hz}, where a
simulation is done to exhibit the void patterns.

Now, suppose that in a narrow $p_T$ window $n_+$ and $n_-$
charged particles are collected in $\Delta\phi$ in azimuthal
angle and $\Delta y = \delta$ in rapidity.  What we suggest is
that $ \left< N_{\rm ch}\right>$ be only about 2 so that $F_{q_1,
q_2}$ can be calculated for $N_{\rm ch} \geq q = q_1 + q_2$ up
to $q$ about 10, as shown in Fig.\,1.  Such values of $N_{\rm
ch}$ are large deviations from $\left< N_{\rm ch}\right>$, but
represent only a tiny fraction of the total multiplicity produced
in a typical event in heavy-ion collisions.  Thus the problem of
overlap from different emission times is minimized, while the
characteristics of charge fluctuations that convey the nature of
the chiral transition are retained.  Of course,  $\left< N_{\rm
ch}\right>$ depends on the sizes of the cuts.  But our
theoretical analysis based on the Ginzburg-Landau formalism
indicates that the scaling exponent $\beta_{q_1, q_2}$ , the
index $\nu$, and the quotient $B$ are all independent of the
window sizes, so long as there is enough statistics to render
accurate determination of $f_{q_1, q_2}$. Since the charge
fluctuations of DCC are much larger than those among the
hadrons produced in the normal critical transition, the
contribution of the latter to our measure is expected to be negligible.

Recently, WA98 presented their data that reveal no correlated
charge-neutral fluctuations in Pb-Pb collisions \cite{na}.
Unfortunately, they have no information on $p_T$ and cannot make the
$p_T$ cut that we regard as essential. The absence of any signature of
DCC in the bulk events does not preclude the possibility of local
condensates of the type we consider here.

It is clear from Eqs.\,(\ref{19}) and (\ref{20}) that $B$
involves order-sum $q$ not greater than 2, whereas the
power-law behavior of $\beta_q$ shown in Fig.\,3 is for $q
> 2$.  Although the former seems simpler to investigate, the
smallness of $q$ requires $\left< N_{\rm ch}\right> \ll 1$ in
order for the factorial moments to exhibit the fluctuation
properties at the far edge of the multiplicity distributions.
That in turn requires the kinematical cuts to be very severe
so that only in rare events can $N_{\rm ch}$ exceed $q$.  As a
consequence, the statistical errors may be large.  For the
latter study of  $\beta_q$, the values of $q$ are larger, so the
cuts can be less severe to allow $\left< N_{\rm ch}\right>$ to be
larger.   The experimental errors on $\beta_q$ then  may
hopefully  be small enough to render the
determination of $\nu$ feasible.  Whether $q$ is small or
large, our analysis probes large fluctuations from the
mean.

Despite the many complications that accompany heavy-ion
collisions, our analysis suggests a possible way of finding the
DCC. If only the bulk events are examined, it is quite possible
that the DCC, even if created, might escape detection. The simple
consideration done here based on the essentials of chiral
transition offers a potentially effective tool to initiate a
first-round experimental exploration.  Since the discovery of a
DCC would provide a definitive confirmation of the accepted
ideas of chiral dynamics in a dense system, such an exploration
is well worth undertaking.

 We wish to acknowledge helpful communications with
Drs.\,Q.\,H.\,Zhang and V.\ Koch.     This work was supported, in
part,  by the U.\ S.\ Department of Energy under Grant No.
DE-FG03-96ER40972.

\vspace*{-1.5cm}
\begin{figure}
\includegraphics[width=0.8\textwidth]{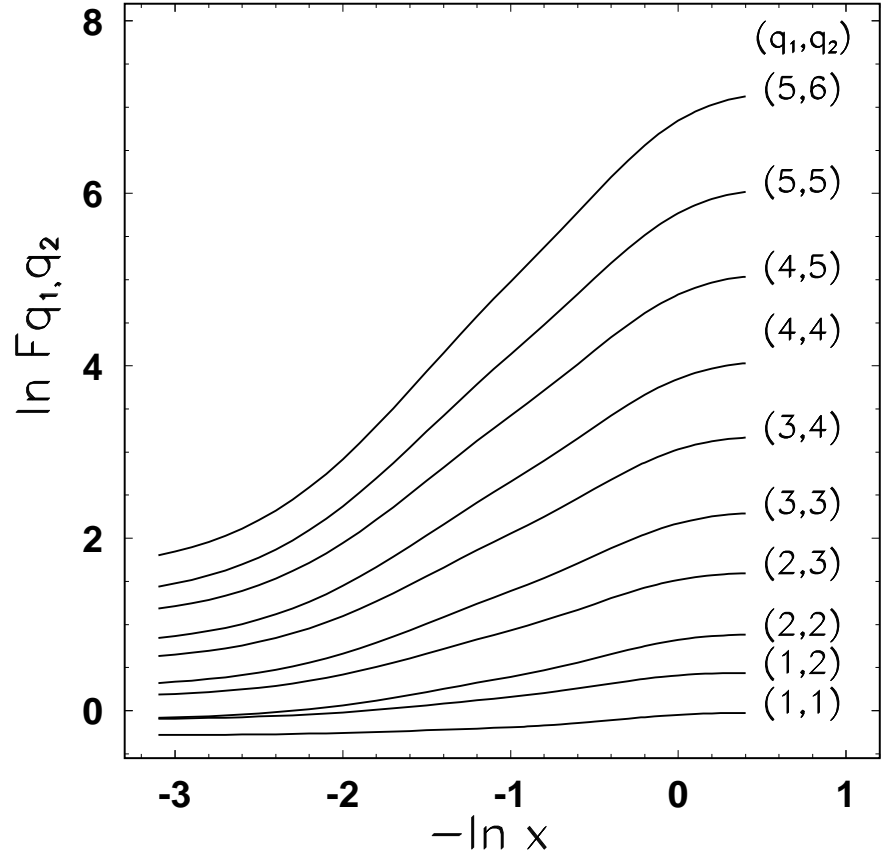}
\caption { \vspace*{-0.3cm}$F_{q_1,q_2}$ vs $x$ for various  orders of the
moments. }
\end{figure}

\vspace*{-1.5cm}
\begin{figure}
\includegraphics[width=0.8\textwidth]{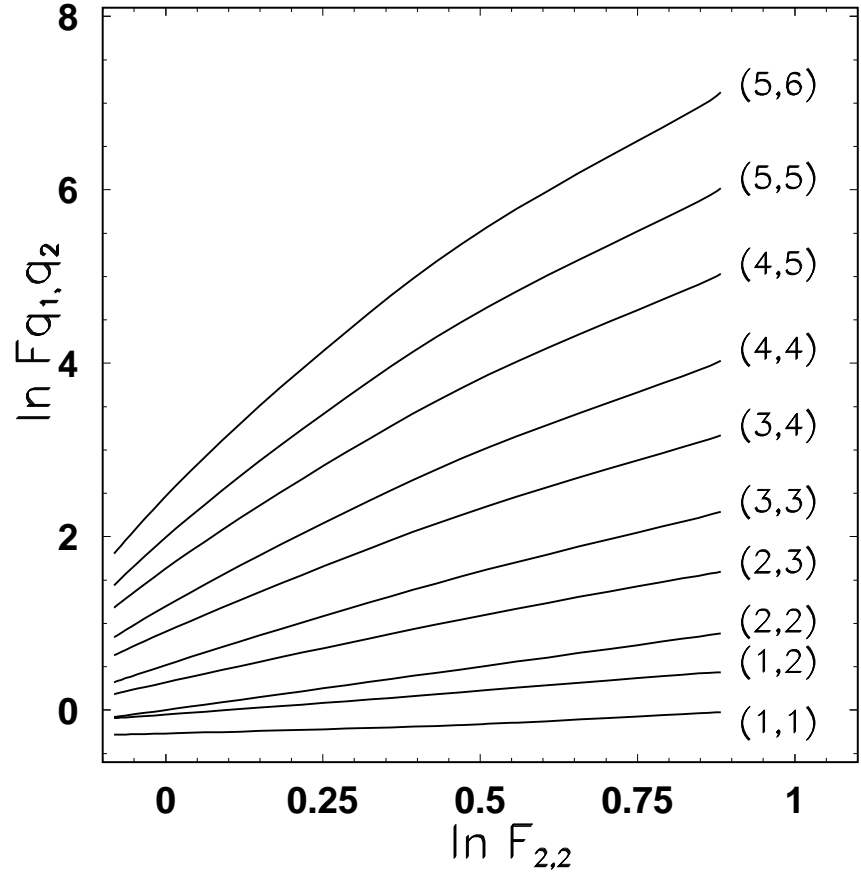}
\caption {\vspace*{-0.3cm}$F_{q_1,q_2}$ vs $F_{2,2}$ in log-log plot.}
\end{figure}

\vspace*{-1.5cm}
\begin{figure}
\includegraphics[width=0.8\textwidth]{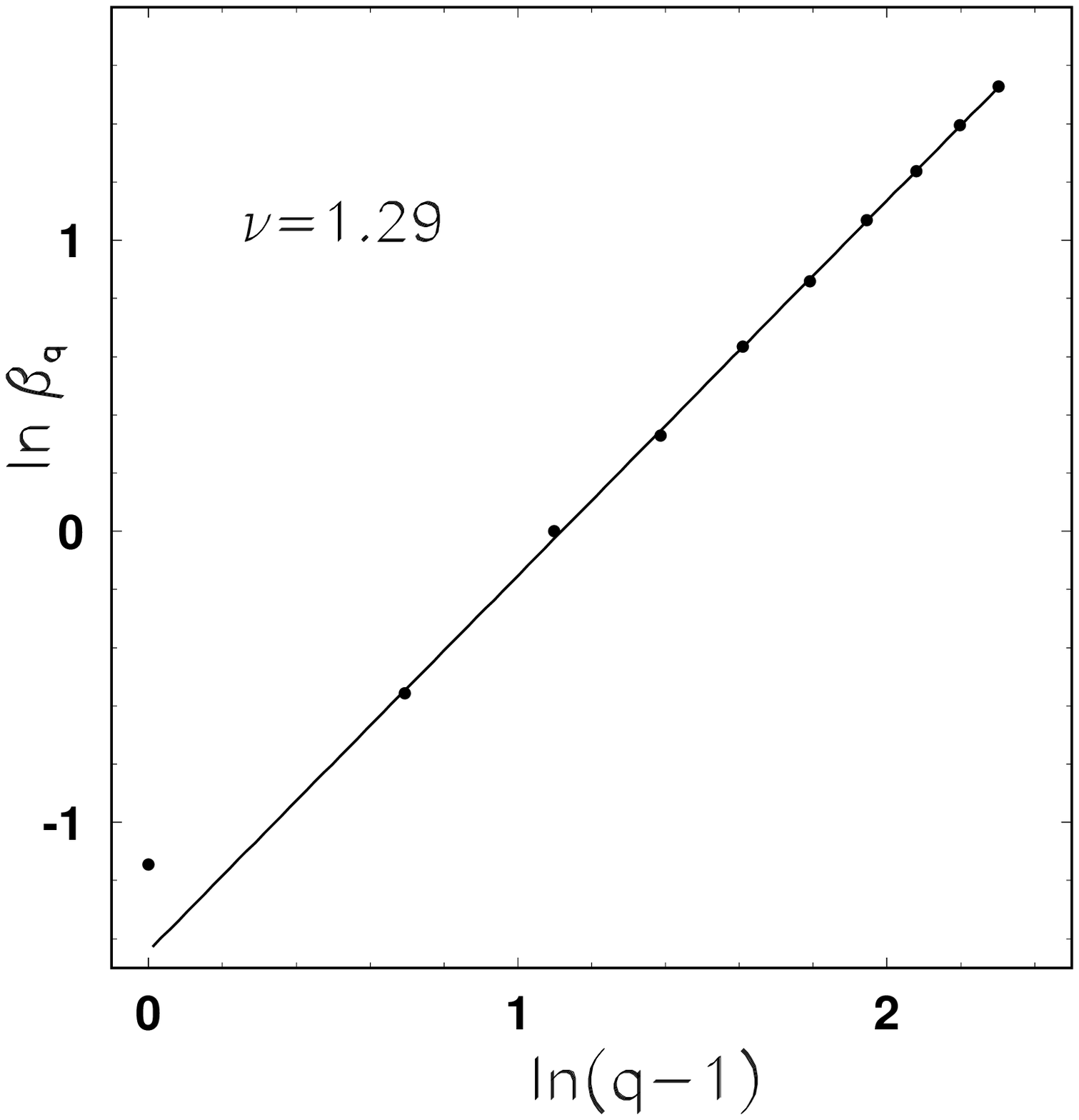}
\caption {\vspace*{-0.3cm}Power-law behavior of $\beta_q$.}
\end{figure}


\begin{thebibliography}{99}

\bibitem{jk} S.\ Jeon and V\ Koch, Phys.\ Rev.\ Lett.\ {\bf
83} (1999) 5435; {\bf 85} (2000) 2076.

\bibitem{ahm}M.\ Asakawa, U.\ Heinz, and B.\ Mueller,
Phys.\ Rev.\ Lett.\ {\bf 85} (2000) 2072.

\bibitem{hj}H.\ Heiselberg, and A.\ D.\ Jackson,
Phys.\ Rev.\ {\bf C 63} (2001) 064904.

\bibitem{gpz} C.\ Gale, V.\ Topor Pop, and Q.\ H.\
Zhang,  McGill preprint (2001).

\bibitem{rev}For a review, see V.\ Koch, talk given at
Quark Matter 2001, 15th Int. Conf. on Ultra-Relativistic
Nucleus-Nucleus Collision, Stony Brook, NY, January 2001.

\bibitem{raj}K.\ Rajagopal, in {\it Quark-Gluon Plasma 2}, edited
by R.C. Hwa (World Scientific, Singapore, 1995).

\bibitem{rw}K.\ Rajagopal and F.\ Wilczek, Nucl.\ Phys.\ {\bf
399} (1993) 395; {\bf 404} (1993) 577.

\bibitem{bk}J.-P.\ Blaizot and A.\ Krzywicki, Phys.\ Rev.\ D {\bf
50} (1994) 442.

\bibitem{hw}Z.\ Huang and X.-N.\ Wang, Phys.\ Rev.\  {\bf D 49} (1994) R4335.

\bibitem{and}I.\ V.\ Andreev, JETP Lett.\ {\bf 33} (1981) 367.

\bibitem{ans}A.\ A.\ Anselm, Phys.\ Lett.\ {\bf B 217} (1988) 169;
A.\ A.\ Anselm and M.\ G.\ Ryskin, Phys.\ Lett.\ {\bf B 266} (1991) 482.

\bibitem{kt}J.\ D.\ Bjorken, Int.\ J.\ Mod.\ Phys.\ {\bf A 7} (1992) 4189.

\bibitem{bj}J.\ D.\ Bjorken, K.\ L.\ Kowalski and C.\ C.\ Taylor,
hep-ph/9309235.

\bibitem{ss}D.\ J.\ Scalapino and R.\ L.\ Sugar, Phys.\
Rev.\ {\bf D 8} (1973) 2284; J.\ C.\ Botke, D.\ J.\
Scalapino and R.\ L.\ Sugar, {\it ibid}.\ {\bf 9} (1974) 813.

\bibitem{gl}V.\ L.\ Ginzburg and L.\ D.\ Landau, Zh.\
Eksp.\ Teor.\ Fiz.\ {\bf 20} (1950) 1064.

\bibitem{hn}R.\ C.\ Hwa and M.\ T.\ Nazirov, Phys.\ Rev.\ Lett.\ {\bf
69} (1992) 741; R.\ C.\ Hwa, Phys.\ Rev.\ {\bf D 47} (1993) 2773.

\bibitem{hp} R.C. Hwa and J. Pan, Phys. Lett. {\bf B 297} (1992) 35.

\bibitem{yc} X.\ Cai, C.\ B.\ Yang, and Z.\ M.\ Zhou, Phys. Rev. 
{\bf C 54} (1996) 2775.

\bibitem{bp} A.\ Bia\l as and R.\ Peschanski, Nucl.\ Phys.\ {\bf
B 273}, 703 (1986); {\bf B 308} (1988) 857.

\bibitem{hz} R.\ C.\ Hwa, and Q.\ H.\ Zhang, Phys. Rev. {\bf C 64} (2001) 054904.

\bibitem{na} T.\ K.\ Nayak (WA98 Collaboration), nucl-ex/0103007.

\end{thebibliography}
\end{document}